%% file: paper.tex
\documentclass[conference]{IEEEtran}
\IEEEoverridecommandlockouts
\usepackage{cite}
\usepackage{amsmath,amssymb,amsfonts}
\usepackage{algorithmic}
\usepackage{graphicx}
\usepackage{textcomp}
\usepackage{xcolor}
\def\BibTeX{{\rm B\kern-.05em{\sc i\kern-.025em b}\kern-.08em
    T\kern-.1667em\lower.7ex\hbox{E}\kern-.125emX}}
\begin{document}

\title{Survey of Attacks and Defenses \\on Edge-Deployed Neural Networks}

\author{
    \IEEEauthorblockN{Mihailo Isakov\IEEEauthorrefmark{1}, Vijay Gadepally\IEEEauthorrefmark{2}, Karen M. Gettings\IEEEauthorrefmark{2}, Michel A. Kinsy\IEEEauthorrefmark{1}}
    \IEEEauthorblockA{\IEEEauthorrefmark{1} Adaptive and Secure Computing Systems (ASCS) Laboratory, Boston, MA
    \\\{mihailo, mkinsy\}@bu.edu}
    \IEEEauthorblockA{\IEEEauthorrefmark{2}MIT Lincoln Laboratory, Boston, MA
    \\\{vijayg, karen.gettings\}@ll.mit.edu}
}

\maketitle

\begin{abstract}
    Deep Neural Network (DNN) workloads are quickly moving from datacenters onto edge devices, for latency, privacy, or energy reasons. While datacenter networks can be protected using conventional cybersecurity measures, edge neural networks bring a host of new security challenges. Unlike classic IoT applications, edge neural networks are typically very compute and memory intensive, their execution is data-independent, and they are robust to noise and faults. Neural network models may be very expensive to develop, and can potentially reveal information about the private data they were trained on, requiring special care in distribution. The hidden states and outputs of the network can also be used in reconstructing user inputs, potentially violating users' privacy. Furthermore, neural networks are vulnerable to adversarial attacks, which may cause misclassifications and violate the integrity of the output. 
These properties add challenges when securing edge-deployed DNNs, requiring new considerations, threat models, priorities, and approaches in securely and privately deploying DNNs to the edge. 
In this work, we cover the landscape of attacks on, and defenses, of neural networks deployed in edge devices and provide a taxonomy of attacks and defenses targeting edge DNNs.
\end{abstract}


\input{introduction}

\input{taxonomy}

\input{attacks}

\input{defenses}

\input{conclusion}

\bibliographystyle{IEEEtran}
\bibliography{paper}

\end{document}

%% file: introduction.tex
\vspace{-0.3cm}
\section{Introduction}
Since the rise of deep learning in the last decade, many different libraries and frameworks for running and training deep neural networks  
(DNN) have been published and open-sourced. In that time, the landscape of software tools for training neural networks has moved from 
difficult-to-install libraries~\cite{caffe}, and support for static graphs only~\cite{theano}, to industry-ready, easy-to-deploy 
frameworks~\cite{tensorflow}, high-development efficiency~\cite{keras}, and support for dynamic graphs and \textit{just-in-time} compilation\cite{pytorch}. 
Recently, as tools have gained maturity, more businesses have started using neural networks in production and exposing services
based on neural networks~\cite{google_nmt}. Deploying neural networks is non-trivial, and the research frameworks proved insufficient 
to handle high-bandwidth, low-latency inference, leading to the development of production-ready frameworks such as Tensorflow 
Serving~\cite{serving} and the standardization of neural network formats with ONNX~\cite{onnx}. With an increasing number of 
mobile devices and PCs possessing GPUs and custom ASICs, networks have been pushed to smartphones~\cite{coreml} and even 
GPU-enabled JavaScript~\cite{tensorflow_js}. With the rise of voice assistants, wearables, and smart cameras, 
the need for low-power inference has led to the development of many custom DNN acceleration ASICs~\cite{edge_tpu, jetson_nano}. 

There are many reasons why businesses or users may want to run neural networks on edge devices, 
as an alternative to sending the data to datacenters for processing, including: 

\noindent \textbf{Privacy:} users may not want or may not be able to send the data to the cloud for privacy or legal reasons. 
For example, a hospital may want to process patient data on servers at a different location, but is not willing to risk patient 
privacy. Even if the patient data is encrypted, if the server is malicious, the patient data may be at risk. 

\noindent \textbf{Power}: sending data directly to the cloud may not be the most power-efficient approach to run neural networks.
For example, in~\cite{7979979}, the authors show that in convolutional neural networks (CNN), processing a few of the first 
convolutional layers before sending the data to the cloud achieves higher power savings compared to processing the whole 
network on the device or sending the input data to the cloud. As more low-power accelerators using approaches such as 
quantization~\cite{NIPS2015_5647}, stochastic computing~\cite{DBLP:journals/corr/RenLLDQQYW16}, or 
sparsity~\cite{DBLP:journals/corr/HanKMHLLXLYWYD16, closnets} are released, we expect the ratio between the cost of processing 
networks and the cost of transmitting input data to become more significant.

\noindent \textbf{Latency:} many applications have hard latency requirements and must process a network within a certain time limit.
Furthermore, for certain mission-critical applications with hard availability guarantees, as in the case of 
autonomous drones or self-driving cars, being able to process data on the device is mandatory.
While datacenters are able to provide virtually unlimited computing power, possibly making inference time negligible, the transmit 
time of inputs over the network often cannot be ignored. Hence, a device must possess the required compute power to process the inputs within the time budget. 

\noindent \textbf{Throughput:} several industries dealing with high bandwidth data are faced with the question of whether to store data for offline processing, allowing thorough analysis at the cost of large amounts of storage, or to process the data in-flight
potentially sacrificing some information, but saving on storage. Take an extreme case: the CERN Large Hadron Collider (LHC) can
generate upwards of hundreds of terabytes of data per second. Storing that data is difficult, so authors of~\cite{lhc} propose 
to process the data in-flight using extremely low-latency FPGA designs.

%% file: taxonomy.tex
\section{Taxonomy of Attacks on and Defenses of Deployed Neural Networks}
Since DNN accelerators have only recently been deployed in commercial products, the field of attacking and defending these devices is in its infancy. 
In this section we aim to provide (1) a taxonomy of DNN accelerator attacks and defenses, and (2) a list of plausible attack surfaces and attacker motivations for targeting edge devices running DNNs.

\subsection{Taxonomy of DNN Accelerator Attacks and Defenses}

\begin{figure*}[t]
    \centering
    \includegraphics[width=0.92\textwidth]{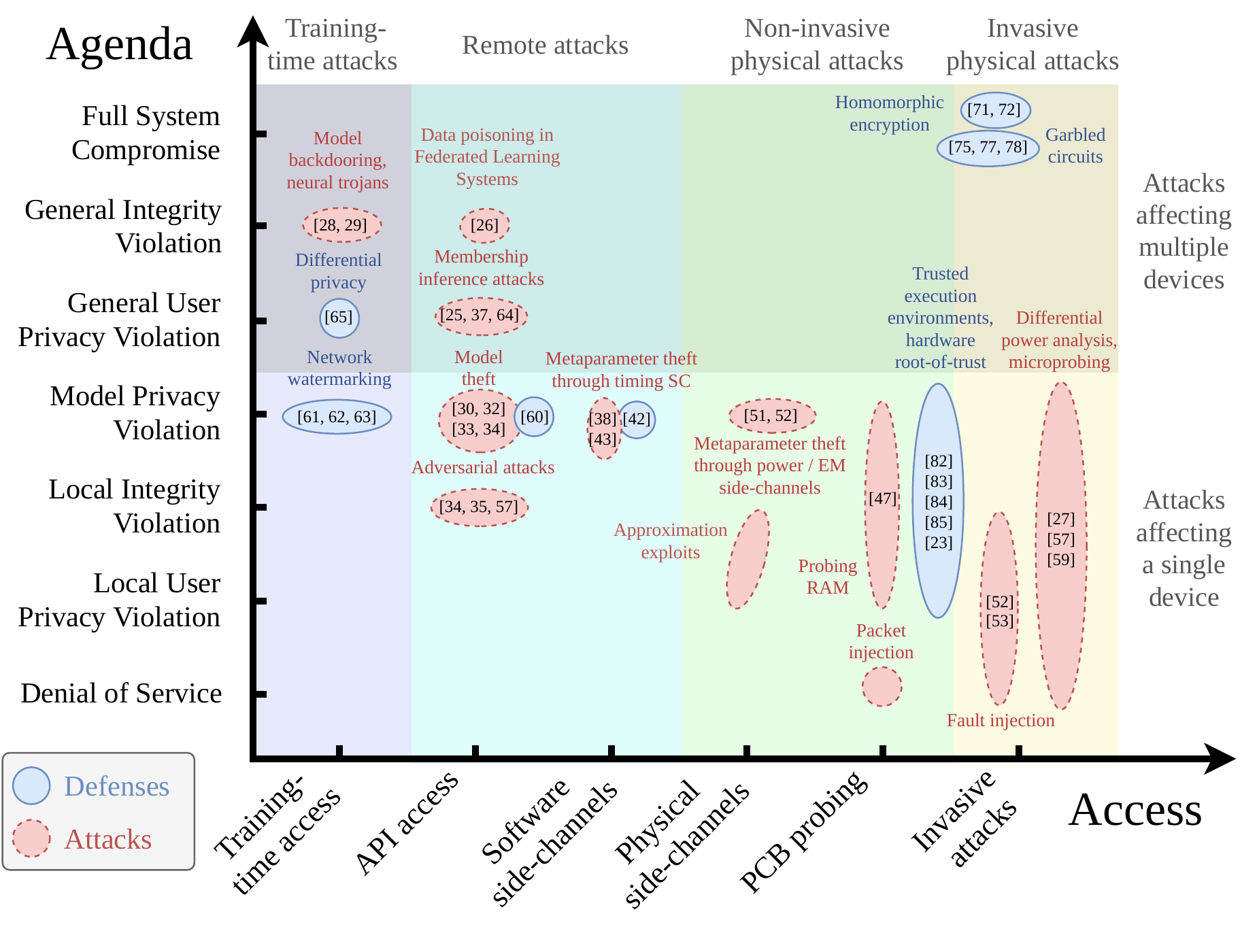}
    \vspace{-0.3cm}
    \caption{Taxonomy of attacks, defenses, and potential vulnerabilities of edge devices running ML inference or on-line training.}
    \label{fig:taxonomy}
    \vspace{-0.4cm}
\end{figure*}

Of the many possible dimensions over which we could characterize attacks and defenses of edge devices, we believe that the \textit{attacker agenda} and \textit{level of access} to the edge device provide a useful classification. The attacker agenda represents the goal of the attacker, and ranges from local denial-of-service (DoS) to gaining full access to a network of edge devices. Level of access is a set of attack surfaces the attacker has access to and ranges from simple API accesses to probing buses or even chip internals. In Figure~\ref{fig:taxonomy}, we present an overview of attacks, defenses, and potential vulnerabilities present in the literature.

\subsection{Attacker Agenda}

The $y$-axis in Figure~\ref{fig:taxonomy} represents the attacker's motivation for attacking an edge-deployed neural network.
We classify attacker motivations into four categories: 

\noindent \textbf{Denial of Service: } attackers may want to prevent a device running a neural network from properly functioning. For example, attackers may want to prevent smart cameras from properly classifying recordings in order not to raise alarms. Denial of Service (DoS) attacks prevent a device from maintaining availability and completing its function. As feedforward neural networks are data-independent and have fixed latencies, DoS attacks targeting DNNs are only applicable to accelerators running data-dependent models, e.g., recurrent neural networks~\cite{LSTM} or neural networks with early exits like BranchyNets~\cite{DBLP:journals/corr/abs-1709-01686} or Tree LSTMs~\cite{tree_lstm}.

\noindent \textbf{User Privacy Violation: } smart devices are increasingly trusted with private user data such as shopping history, voice commands, or medical recordings~\cite{DBLP:journals/corr/abs-1711-05225}. This data is valuable for its advertising, monitoring, or polling value. User privacy violations are cases where the attacker is able to access measured or stored sensor data from the device or user data the device from the network. For example, attacks on voice assistants where the attacker can access previous voice commands constitute a local privacy violation. 

\noindent \textbf{Model Privacy Violation: } the attacker may attempt to exfiltrate a neural network model for a number of reasons: (1) models require significant investment to develop, and, as such, may be stolen and sold, or used in ensembles as a black box~\cite{tie}, (2) finding adversarial examples is significantly easier if the attacker has access to a model (i.e., the white-box scenario), compared to only having access to model inputs and outputs (i.e., the black-box scenario)~\cite{DBLP:journals/corr/PapernotMG16}, or (3) the attacker may attempt to learn data from the dataset the model was trained on~\cite{DBLP:journals/corr/abs-1802-08232}. 

\noindent \textbf{Integrity Violation: } the attackers may not want to outright prevent the device from functioning, but may want to force the neural network to perform in an unacceptable way. For example, malware may craft adversarial packets in an attempt to fool a network intrusion detection system (IDS) that uses DNNs to identify packets. Local integrity violations are cases where the attacker is able to affect the correctness of a device's neural network inference.

In Figure~\ref{fig:taxonomy}, we list several attacker agendas, sorted by severity. We add two additional categories of general user and integrity violations, which consists of cases that affect not only a single device, but multiple devices, some of which are not under the attacker's physical control. An example of this is data poisoning attacks on federated learning systems~\cite{Bagdasaryan2018HowTB}, where attackers controlling one device can insert backdoors into all devices in the network.

\subsection{Attacker's Level of Access}
The $x$-axis in Figure~\ref{fig:taxonomy} represents the attacker's level of access to an edge device. The five access categories vary by invasiveness from purely software, API-based attacks and defenses, all the way to invasive attacks, such as decapsulation and microprobing. 
The API attacks assume that the attacker only has access to the device through conventional channels, e.g. through the network (as in the case of machine translation systems) or the device's sensors (as in the case of voice assistants). Software side-channels additionally assume that the attacker has some ability to measure side-channels through the device's legitimate outputs, e.g., measure the latency of network responses or the amount of traffic the device is sending to the cloud. For both API and software side-channel attacks, the attacker does not need physical contact with the device. Additionally, these attacks are typically simple to automate, unlike the attacks based on physical properties of the device. In the case of physical side-channels, the attacker needs physical proximity to the device, as in the case of power or electromagnetic (EM) analysis. However, physical side-channel attacks do not require invasive sensors or access to the printed circuit board (PCB). PCB probing attacks include attacks that may measure data or timing information of any bus exposed on the PCB, but not the internals of any chip on the device. These attacks include probing RAM or non-volatile memory (NVM), as well as cold-boot attacks, etc. 
Finally, invasive attacks access the internals of a chip. These include approaches such as decapsulation (a procedure where the chip packaging is removed), microprobing (where the attacker can probe the internals of a chip), chemical attacks that can reveal information stored in read-only memory, and scanning electron microscope (SEM) attacks (which are able to read RAM memory)~\cite{Tria2011}. These attacks typically require specialized labs and expensive equipment. They are often destructive and may require multiple devices before a successful attack is implemented. For completeness, we also include training-time attacks and defenses that happen before devices are deployed, or during on-line training. 
Attacks that take place strictly before deployment are beyond the scope of this study.

%% file: attacks.tex
\section{Attacks on Deployed Neural Networks}
We present a short survey of published attacks on neural network accelerators. We focus primarily on test-time attacks (attacks on already trained models), as we assume that training-time attacks such as data poisoning~\cite{Biggio2012PoisoningAA} or Neural Trojans~\cite{Trojannn} must happen before the model is deployed. 

\noindent \textbf{API attacks:}
API attacks interact with the victim device only through the sensors, the interface, or the network. 
Here we assume that the attack is independent of the hardware platform running the neural network and does not rely on any side-channel information.
The majority of the API attacks present in the literature either attempt to (1) exfiltrate the model or the model metaparameters, (2) find adversarial examples, or (3) infer some property of the model's training data.

In~\cite{stealing_nns}, the authors show how machine learning models hosted behind APIs can be exfiltrated. Here the attacker sends crafted inputs and collects outputs from the model until the attacker is able to reconstruct the model behind the API. In the case of simpler ML models such as decision trees, the models can be perfectly reconstructed. However, for more complex models such as neural networks, the attacker cannot simply solve nonlinear equations to arrive at model weights, but must instead train a `student' network on input-output pairs collected from the API~\cite{2015arXiv150302531H}. A similar work~\cite{2017arXiv171101768O} shows the simplicity of reverse-engineering black-box neural network weights, architecture, optimization method and the training/data split. In~\cite{Orekondy}, authors reframe the goal from model theft, to arriving at a `knockoff' model exhibiting the same functionality.  In~\cite{metaparameter_theft}, authors ignore model parameters and instead attempt to steal the hyperparameters of a network. Good hyperparameters, while far smaller than models, can be more difficult to arrive at, as they require many experiments and human effort to tune. 

In order to violate the integrity of a machine learning model, attackers may attempt to find adversarial examples~\cite{2014arXiv1412.6572G}. While most attacks rely on having access to the white-box model or the output gradient, several works have shown that even black-box networks~\cite{2017arXiv171101768O} and networks with obfuscated gradients~\cite{DBLP:journals/corr/abs-1802-00420} are not resistant to determined attackers.

Lastly, attackers may attempt to infer some information about the data the neural network was trained on. Attacks which determine whether a specific input was used in training a model are called \textit{membership inference} attacks. Though DNN models are typically smaller than the training dataset, they can nonetheless memorize potentially secret information~\cite{DBLP:journals/corr/abs-1802-08232}, as in the example of predictive keyboards memorizing PIN codes or passwords. In~\cite{Salem2018}, authors show that even when the model is behind a black-box API, and the adversary has no knowledge of the victim's training dataset, membership inference attacks are still successful.

\noindent \textbf{Software side-channel attacks:}
API and software side-channel (SC) attacks target a similar attack surface, but software side-channel attacks can additionally gain information through side-effects such as timing or cache side-channels. Here, the attacker abuses information about the physical device processing the attackers request to gain an insight into the internal state of the device. 

Both timing and cache side-channels typically cannot reveal anything about the data being processed on the device - timing SC reveal information about the compute intensity of a certain task, and cache SC reveal information about recently accessed addresses in the caches. As such, they are commonly employed to extract course-grain information such as neural network architecture running on a device. For example, in Cache Telepathy~\cite{Yan2018}, attackers use the Flush+Reload~\cite{184415} and Prime+Probe~\cite{Liu:2015:LCS:2867539.2867673} cache SC attacks to measure the size of general matrix multiply (GEMM) operations, first counting the number of parameters in the model, and then narrowing down the model architecture. While this attack is restricted to CPUs, GPUs are no less vulnerable to cache-side channels~\cite{Naghibijouybari2018}. A similar work~\cite{DBLP:journals/corr/abs-1810-03487} is applicable to CPUs, GPUs, and DNN accelerators, and can fingerprint a network after only a single inference operation. It leverages a priori knowledge of major DNN libraries to prime the instruction cache and learn which functions are called during inference.

Timing attacks are also used to reveal model architecture: in~\cite{nn_timing_attack}, the authors assume that the attacker knows the victim's hardware, and is able to buy the same device in order to build timing profiles of different networks. By only knowing the accuracy and the latency of the victim network, the attacker trains many candidate architectures searching for one that has the same signature. This, however, requires the attacker to first steal a part of the training dataset using a membership inference attack~\cite{Salem2018}, which negates much of the need for stealing a model architecture.

Software SC attacks may be less successful in the edge domain compared to the cloud, as edge devices typically serve a single user, while SC are typically used for compromising secure multi-user systems~\cite{cross_vm}. However, as more networks are pushed to the edge, we can expect multi-network systems with different privileges, goals, and timescales to become increasingly common. An example of this may be predictive keyboards, which perform both inference (text prediction) and NN training on the same device~\cite{DBLP:journals/corr/McMahanMRA16}.

Another potential vulnerability may be introduced with the adoption of data-dependent inference latency. For example, DARPA's N-ZERO program~\cite{darpa} seeks low-power edge devices that may need to stay dormant for years and have several levels of neural networks, each activating the next one once a certain pattern is sensed. These types of networks are inherently vulnerable to timing attacks, as conventional methods for defending against timing attacks, such as constant time functions negate all the benefits of variable-latency inference. 

\noindent \textbf{Physical side-channel attacks:}
Physical side-channels typically measure some physical quantity, such as power, electromagnetic radiation, vibration, etc.
Several works have explored using physical side-channels to extract the neural network architecture, weights, or user inputs to an edge device.

Memory access patterns can trivially reveal model architecture. In~\cite{Hua2018}, authors attempt to steal a model and model architecture running on a secure enclave such as Intel SGX~\cite{Costan2016IntelSE}, by observing memory access traces. While traces allow attackers to learn the architecture, the model can only be stolen if the accelerator exploits data-dependent model properties, such as the sparsity of hidden neuron activations~\cite{cnvlutin}. Power and electromagnetic (EM) side-channel attacks are explored in~\cite{Batina}, where the authors use EM SCs to learn the activation function, simple power analysis to learn model architecture, and differential power analysis to learn network weights. While simple power analysis does not require invasive measures, differential power analysis may require chip decapsulation, and would need to be classified as an invasive attack. Finally, the authors show how user's private inputs may be extracted using power analysis.
A similar attack is explored in~\cite{DBLP:journals/corr/abs-1803-05847}, where the authors use a power side-channel to observe the processing of the first layer of a convolutional network and extract user's inputs. The authors explore both active and passive attackers, i.e., attackers that can actively input their own images to the accelerator and attackers that can only observe user inputs.
Another line of attacks attempts to induce faults in order to cause misclasifications~\cite{Liu, Rakin} and relies on a microarchitectural or device-level attacks, such as RowHammer~\cite{Gruss2017}.

\noindent \textbf{Probing attacks:}
Probing attacks assume that an attacker is able to access the individual components of the device, e.g., the CPU/GPU/ASIC, the RAM memory, non-volatile storage, or busses, but is not able to perform invasive attacks that access the internals of the chips. The attacker has full access to measure signals on any exposed wires or even drive wires themself. This opens up a variety of denial-of-service, integrity, and privacy attacks. Additionally, probing attacks assume that no tamper evidence is left after the attack, unlike invasive attacks. 

A simple attack the attacker can carry out is model theft - here the attacker probes the memory bus and runs an inference operation while recording the model being loaded onto the chip. This can be prevented by storing only the encrypted model in RAM and NVM, and decrypting the model on-the-fly, if power requirements permit~\cite{tie}. 
However, even if the model is encrypted, just knowing the memory access pattern is enough to reveal the model architecture. Each layer and activation will have a different memory bandwidth, and the attacker can monitor these changes along with memory addresses to learn where layers start and end in memory. While oblivious RAM~\cite{oram} can hide memory addresses, memory access timings are still sufficient to reveal the topology of the model. This forces the defender to either prefetch weights or create fake accesses in order to obfuscate memory access timings~\cite{tie}.
Similarly, network activations may be larger than the available on-chip memory and may be stored in RAM. These activations also need to be encrypted, because even in cases when the device manufacturer is not concerned about privacy, these activations can be used in order to infer the model weights~\cite{Milli}. 

The attacker may also attempt to overwrite parts of RAM or feed their own inputs to the chip in order to subvert any software guards, for example in order to generate more input-output pairs used for API model theft~\cite{stealing_nns}. Encrypted RAM may defend against this type of attack, but the device is still susceptible to DoS attacks, where fake accesses are inserted on busses. 

\noindent \textbf{Invasive attacks:}
Invasive attacks assume that the attacker has full control over the chip and is able to bypass any tamper-proof packaging.
These attacks include freezing the device in order to extract volatile memory, probing the internals of the chip, ionizing parts of the chip in order to induce faults, feeding non-legitimate voltages and clock frequencies to the chip, etc. Mounting these attacks is typically cost-prohibitive and requires substantial expertise and equipment to execute. 

Several works have explored invasive attacks on DNN accelerators, and many of the conventional (non-DNN specific) invasive attacks are still applicable to them. In DeepLaser~\cite{Breier}, the authors decapsulate a chip and are able to induce faults by shining a laser on the chip, causing misclasifications by the neural network. This is done by causing bit-flips in the last layer's activation function, where flipping high-order bits of an output neuron's activation will cause the associated category or value to be dominant. Choosing the minimal amount of bit-flips to achieve a desired output has been studied in two works: \cite{Liu} and \cite{Rakin}. Both these works show that, despite the robustness of neural networks to random perturbations, networks are highly susceptible to targeted bit-flips, in a manner similar to non-targeted adversarial attacks~\cite{Yuan}.

While we have not been able to find any examples of this, we expect neural network accelerators to be vulnerable to cold boot attacks~\cite{Halderman:2009:LWR:1506409.1506429}, which may be able to steal unencrypted models stored in volatile memory, or microprobing~\cite{Tria2011}, which may be able to bypass model or user data decryption.

%% file: defenses.tex
\section{Defending Edge Devices Running Neural Networks}
We briefly cover proposed defenses for edge devices running neural networks. 

\noindent \textbf{API defenses:}
The majority of API attacks we have mentioned attempt to steal the model or the model architecture, learn which inputs have been used to train the model, or find adversarial examples for the model running on the device. As finding adversarial examples typically involves first stealing the model~\cite{2017arXiv171101768O}, we focus only on defenses against model exfiltration and membership inference attacks.

In a recent work called Prada~\cite{Juuti}, the authors succeed in detecting API model-stealing attacks with a 100\% detection rate and no false positives. Here, the authors do not attempt to detect if a single query is malicious (as in the case of adversarial attacks), but whether some consecutive set of them is actively trying to steal the model. The authors detect model-stealing queries as they are specifically crafted to extract the maximum amount of information out of the model. However, the authors note that attackers may introduce dummy queries to maintain a benign query distribution, resulting in slower but more covert model-stealing attacks.

Watermarking is a method for embedding secret information into some system in order to verify the origin of that system at a later date. Watermarking has been proposed as a method of establishing ownership of neural networks~\cite{Adi2018, Uchida, DBLP:journals/corr/abs-1804-00750}. Here, a watermark is applied to a neural network in such a way that it does not impact the network's accuracy, but can be used to confirm ownership from network outputs. Even if the party responsible for the theft attempts to prune or finetune the network, watermarks can be retained~\cite{Uchida}.

Defending against membership inference attacks has been explored in several works. In~\cite{Shokri2017}, the authors claim that overfitting is the reason why models are vulnerable to membership inference attacks and suggest that differential privacy~\cite{Abadi2016} used during training can protect against these types of attacks. They propose several defenses, similar to those used in defending against adversarial attacks: (1) reducing the number of predicted classes (in the case of classification problems), (2) reducing the amount of information per class by rounding prediction probabilities, (3) increasing entropy of the prediction values and (4) using stronger regularization during training.
Similarly, in~\cite{Salem2018}, the authors propose two defenses: dropout~\cite{JMLR:v15:srivastava14a}, where authors show that randomly zeroing out neurons during training partially prevents the attackers from inferring membership, and model stacking, where multiple models are used in an ensemble to make a prediction. 

\noindent \textbf{Side-channel defenses:}
Due to the data-independent behavior of non-recurrent DNNs, all of the software side-channel attacks we have listed attempt to steal the network architecture. We have not been able to find any attacks that succeed at violating privacy of the inputs or the model parameters through software side-channels.
In DeepRecon~\cite{DBLP:journals/corr/abs-1810-03487}, where attackers prime the instruction cache in order to learn function invocations, the authors propose a defense where the defender simultaneously creates decoy function calls to similar neural network layers. These decoy layers should be small enough not to incur a performance penalty. However, this defense does not stop the attacker from using data cache-based side-channels or timing side-channels.
Cache Telepathy~\cite{Yan2018} suggests less aggressive compiler optimizations, cache partitioning~\cite{7446082} or disallowing resource sharing as defenses against cache-based SC. However, these may not be viable solutions without hardware support for secure caches.

While cache-based defenses may help hide some of the accesses, and the defender may go so far as to remove the possibility of an attacker executing code on the same shared resources as the victim, a determined attacker may attempt to probe the memory bus. As neural networks are typically larger than the last-level cache of modern processors, caches will suffer from capacity misses and the network architecture may be exposed to memory probing attacks.
In the Trusted Inference Engine (TIE)~\cite{tie}, the device can either create fake memory accesses in times of reduced memory bandwidth or prefetch data, given available on-chip storage. As TIE targets networks with data-independent profiles (i.e., not recurrent neural networks), the timing of fake or prefetched accesses can be calculated at compile time.

Similar techniques can be applied to counter power and timing side-channels. As long as networks have data-independent behavior, i.e., the accelerator does not attempt to take advantage of zero values~\cite{cnvlutin}, or the network computation graph is static~\cite{DBLP:journals/corr/SutskeverVL14, DBLP:journals/corr/abs-1709-01686}, power and timing side-channel attacks should not be able to learn information about the network. 

\noindent \textbf{Defenses against invasive and semi-invasive attacks:}
There are two common approaches used when an organization needs to deploy software with privacy or integrity requirements. One option is to not trust the edge hardware, and assume that the hardware can be actively malicious, as in the case of untrusted CPUs/GPUs, possible hardware Trojans, broken hardware defenses~\cite{206170}, etc. 
There exist several algorithms that allow processing on private data. Homomorphic encryption~\cite{Rivest1978} (HE) for neural networks has been explored in CryptoNets~\cite{cryptonets}, where the authors use HE to run neural networks on encrypted data, without decrypting it at any time during the process. One of the issues with using HE is the performance reduction - inference using HE can be 100-1000 times slower than without HE. Several works have, however, been able to accelerate HE for neural networks. In Gazelle~\cite{Juvekar}, authors leverage HE for linear layers and Yao's Garbled Circuits~\cite{Goldreich:2003:CCP:966037.966044} for offloading calculating nonlinearities to the owner of the private data, as well as an efficient SIMD implementation and a set of homomorphic linear algebra.

While HE is very efficient for linear layers of a network, DNNs typically use nonlinear activations between the layers, requiring many rounds of computationally expensive calculations. An alternative venue for private inference is based on Yao's Garbled Circuits~\cite{Goldreich:2003:CCP:966037.966044} (GC). Here, two parties want to compute the output of a function (a neural network in this case), where one party supplies the network, and the other the inputs to the network. The party that supplies the network typically creates a garbled circuit and uses a procedure such as oblivious transfer~\cite{Even:1985:RPS:3812.3818} (OT) to acquire the second party's inputs without learning those inputs. A naive implementation of neural networks on GC is very inefficient, and several works have presented domain-specific optimizations to them. In DeepSecure~\cite{Darvish2018}, authors first prune the network~\cite{Han2016}, and then convert the network to Verilog for which they can apply logic minimization. In~\cite{Ball2019}, authors present a modified GC that supports free addition and constant-multiplication on a limited integer range, and a significantly cheaper activation function.  
As a third take on efficient DNNs using GC, XONN~\cite{xonn} attempts to accelerate XNOR-based networks~\cite{DBLP:journals/corr/RastegariORF16} (networks where activations have only values of -1 or 1), as XNOR operations can be processed for free in GC~\cite{10.1007/978-3-540-70583-3_40}. While GC requires a linear number of rounds w.r.t. the number of network layers, both~\cite{Darvish2018} and~\cite{xonn} are able to perform inference in a fixed amount of rounds.

The question that arises is whether it makes sense to run any of these algorithms on edge devices. In the case of inference, where both the model and user inputs should be kept private, the defender has the choice of sending encrypted inputs to the cloud or sending the encrypted model to the edge. Since HE is computationally expensive, edge devices may not receive any latency benefits by running the models locally (unless they are not connected to the network at all). 

Another option for private edge inference is \textbf{\textit{hardware root-of-trust}}~\cite{Tehranipoor:2011:IHS:2051742}. Here, the defender trusts some type of hardware device, which is built with certain security measures, as in the case of secure enclaves~\cite{Costan2016IntelSE, Suh:2003:AAT:2591635.2667184, keystone} or secure accelerators~\cite{tie}. These devices are typically built to work in adversarial environments, where the threat model assumes that attacker can tamper with the device, but cannot probe chip internals.
For example, using secure enclaves, such as Intel SGX~\cite{Costan2016IntelSE}, to perform inference can provide privacy and integrity to the user and neural network deployer, but may be very inefficient. In MLCapsule~\cite{MLCapsule}, authors develop a machine learning as a service (MLaaS) platform above Trusted Execution Environments (TEE) such as Intel SGX, and formally prove it's security. In~\cite{Tramer2018}, the authors propose to use Intel SGX as a hardware root-of-trust, but leverage other hardware such as more powerful (but untrusted) CPUs cores and GPUs to perform inference. The authors are able to guarantee both the privacy of the data sent to untrusted devices, as well as the integrity of results received.
An alternative venue explores building custom secure neural network accelerators~\cite{tie}. Here, the design stores obfuscated or encrypted models in off-chip memory, and performs efficient decryption / deobfuscation on the device. The design leverages secure pseudo-random number generators using physical unclonable functions~\cite{4261134} (PUF) as a source of randomness as an alternative to the power-hungry but more secure encryption. The design also provides security against timing attacks by prefetching data or creating fake accesses to RAM memory. 

Since the attacker can still probe peripherals, the device must encrypt data in RAM. However, by timing the memory accesses, the attacker can learn the model architecture. Using oblivious RAM (ORAM) does not help, as ORAM only protects the address values and not access times. Additionally, neural network weights are typically stored in ascending order, so knowing the addresses (but not timings) reveals only the complete model size. To prevent the attacker from timing the RAM, the defender, then, must either have a prefetcher and load weights in advance while maintaining a constant bandwidth, or create fake accesses in times when the bandwidth is unused~\cite{Fletcher:2012:SPA:2382536.2382540, tie}.

%% file: conclusion.tex
\vspace{-0.1in}
\section{Conclusion}
In this work, we have presented a survey of attacks and defenses on neural networks. We have created a taxonomy of attacks and defenses with regard to attackers level of access to the hardware, and attacker's agenda. We have described different types of attacks on neural networks, ranging from API-based attacks to invasive attacks such as decapsulation and microprobing. Finally, we gave an overview of the types of defenses of neural networks, with the goal of protecting the privacy of user data, the privacy of deployed neural networks, or the integrity of neural network inference.